\documentclass[twocolumn,pra,aps,showpacs]{revtex4-1}
\usepackage{xspace,amsmath,amsfonts,amsthm,amssymb,amsbsy,graphicx,color}

\begin{document}

\newtheorem{theo}{Theorem} 
\newtheorem{lemma}{Lemma}

\title{A wave-function Monte Carlo method for simulating conditional master equations}

\author{Kurt Jacobs} 

\affiliation{Department of Physics, University of Massachusetts at Boston, Boston, MA 02125, USA}

\affiliation{Hearne Institute for Theoretical Physics, Louisiana State University, Baton Rouge, LA 70803, USA}

\begin{abstract}
Wave-function Monte Carlo methods are an important tool for simulating quantum systems, but the standard method cannot be used to simulate decoherence in continuously measured systems. Here we present a new Monte Carlo method for such systems. This was used to perform the simulations of a continuously measured nano-resonator in [Phys. Rev. Lett. {\bf 102}, 057208 (2009)].  
\end{abstract}


\maketitle 

\section{Introduction}
The now standard ``wave-function Monte Carlo method" for simulating the evolution of a quantum system undergoing decoherence is a very important numerical tool~\cite{Dalibard92, Hegerfeldt92, Molmer93, Wiseman01, Jacobs09b}. This method allows a simulation of the density matrix, an object of size $N^2$ where $N$ is the dimension of the system, to be replaced by a simulation of a number of pure states, each of which is only of size $N$. With the increasing relevance of continuous measurement~\cite{JacobsSteck06} and feedback control~\cite{Jacobs08b} to experimental quantum systems, especially in superconducting circuits~\cite{Schuster07, Houck07} and nanomechanics~\cite{Blencowe04, Hopkins03, Naik06, Jacobs07b, Regal08}, one needs to simulate continuously measured systems subject to decoherence. The standard Monte-Carlo method cannot be used in this case, because it applies to master equations but not to {\em stochastic} (or conditional) master equations (SME's). 

To date, two Monte Carlo methods have been devised for simulating conditional master equations, but both suffer limitations. The first is by Gambetta and Wiseman~\cite{Gambetta05}, who used the linear formulation~\cite{GG,Wiseman96,JacobsSteck06} of quantum trajectories to derive their method. The less desirable feature of this method is that it requires evolving a fraction of ensemble members that end up contributing negligibly to the final density matrix, and to this extent it is inefficient. The second method, recently suggested by Hush et al.~\cite{Hush09}, is specifically designed for simulating systems with very large state-spaces, in which it is not possible to use wave-function methods. This requires the use of a quasi-probablity density, such as the Wigner function, and is therefore not as simple to apply to many systems. Further, the elements in the ensemble for this method are not wave-functions but points in phase space. This is important for very large state-spaces, but less desirable when wave-functions (pure-states) can be used. Here we present a wave-function Monte Carlo method that avoids all the above issues.  This method was used to perform the simulations in reference~\cite{Jacobs09}, but the details were not presented there. 

In the next section we state the standard Monte Carlo method for reference purposes. In section~\ref{sec::meth} we present the new method with a minimum of discussion. The purpose is that this section should serve as an easily accessible reference for anyone wanting to implement the method. We also note that a parallel implementation using C++/MPI is available from the author's website~\cite{Website}. In section~\ref{sec::deriv} we show how the method is derived, and thus show that it reproduces the evolution of a stochastic master equation. In section~\ref{sec::num} we use the method to simulate a measurement of the energy of a harmonic oscillator, and compare it to a direct simulation of the SME. Section~\ref{sec::conc} concludes with a summary of the results. 

\section{The Standard Monte Carlo Method}
In what follows, $L$ and $M$ are operators, $\rho$ is the density matrix, and $dW$ is a Wiener process, independent of any other Wiener processes that may be introduced. 

The standard wave-function Monte Carlo method is implemented as follows(see, e.g.~\cite{Wiseman01}). To simulate the master equation 
\begin{eqnarray}
 \dot{\rho} & = & - \gamma ( L^\dagger L \rho + \rho L^\dagger L - 2 L \rho L^\dagger ) 
\label{ME}
\end{eqnarray}
we perform the following steps: 

1. Create a set of $N$ pure states $|\psi_n\rangle$, so that the desired initial value of $\rho$ is approximately  
\begin{eqnarray}
  \rho(0) =  \frac{1}{N} \sum_{n=1}^N |\psi_n\rangle \langle\psi_n | . 
\end{eqnarray}

2. Evolve each pure state by repeating the following steps (i and ii): 

i) Increment each state using the stochastic Schr\"{o}diner Equation (SSE)  
\begin{eqnarray}
 d |\psi_n\rangle & = & - \gamma  \left[ L^\dagger - 2 \left\langle L + L^\dagger \right\rangle_n \right] L |\psi_n\rangle dt \nonumber \\ 
 & &  +  \sqrt{2\gamma} L  |\psi_n\rangle dV_n , 
\end{eqnarray}
where 
\begin{equation}
   \left\langle L + L^\dagger \right\rangle_n \equiv \langle \psi_n |(L + L^\dagger)| \psi_n \rangle , 
\end{equation}
and the $dV_n$ are mutually independent Wiener noise increments satisfying $(dV_n)^2 = dt$. 

ii) Normalize each of the $|\psi_n\rangle$. 

3. The density matrix at time $t$ is (approximately) 
\begin{eqnarray}
  \rho(t) =  \frac{1}{N} \sum_{n=1}^N |\psi_n(t)\rangle \langle\psi_n(t) | . 
\end{eqnarray}

\section{The new Monte Carlo method} 
\label{sec::meth}

The conditional (stochastic) master equation  
\begin{eqnarray}
 d\rho & = & - \gamma ( L^\dagger L \rho + \rho L^\dagger L - 2 L \rho L^\dagger )dt \nonumber \\ 
  & & - k ( M^\dagger M \rho + \rho M^\dagger M- 2 M \rho M^\dagger )dt \nonumber \\ 
  & & + \sqrt{2k}(M\rho +\rho M^\dagger - \langle M + M^\dagger \rangle \rho )dW 
\label{SME}
\end{eqnarray}
describes a measurement of $M$, and decoherence due to an interaction with $L$. 

To simulate the above SME we perform the following steps: 

1. Create a set of $N$ pure states, and $N$ probabilities $P_n$, so that the desired initial value of $\rho$ is approximately 
\begin{eqnarray}
  \rho(0) =   \sum_{n=1}^N P_n |\psi_n\rangle \langle\psi_n | ,  \;\;\;\;\; \sum_{n=1}^N P_n = 1. 
\end{eqnarray}
Since the $P_n$ are the weightings of the pure states in the ensemble that forms $\rho$, the effective size of the ensemble is no longer $N$, but can be characterized, for example, by the exponential of the von Neumann entropy of the set $\{P_n\}$: 
\begin{eqnarray}
  N_{\mbox{\scriptsize eff}} =   \exp \left[ - \sum_{n=1}^N P_n \ln P_n  \right]  \leq N . 
\end{eqnarray}
This effective size is maximized (equal to $N$) iff all the $P_n$ are equal to $1/N$. We therefore choose $P_n = 1/N$ as the initial values of the weightings. 

2. Evolve each pure state by repeating the following steps (i -- vii):  

i) Increment each state using the SSE  
\begin{eqnarray}
 d |\psi_n\rangle & = & - \gamma  \left[ L^\dagger - 2\left\langle L + L^\dagger \right\rangle_n \right] L |\psi_n\rangle dt \nonumber \\ 
 & &  +  \sqrt{2\gamma} L  |\psi_n\rangle dV_n , 
\end{eqnarray}
where 
\begin{equation}
   \left\langle L + L^\dagger \right\rangle_n \equiv \langle \psi_n |(L + L^\dagger)| \psi_n \rangle , 
\end{equation}
and the $dV_n$ are mutually independent Wiener noise increments satisfying $(dV_n)^2 = dt$. 

ii) Normalize each of the $|\psi_n\rangle$.  

iii) Increment each state by 
\begin{eqnarray}
 d |\psi_n\rangle & = & - \gamma  \left[ M^\dagger - 2\langle M + M^\dagger \rangle \right] M |\psi_n\rangle dt \nonumber \\ 
     & &  +  \sqrt{2\gamma} M  |\psi_n\rangle dW , 
\end{eqnarray} 
where 
\begin{equation}
   \langle M + M^\dagger \rangle \equiv  \sum_{n=1}^N P_n \langle \psi_n |(M + M^\dagger)| \psi_n \rangle . 
\end{equation}

iv) Update the probabilities $P_n$ using 
\begin{equation}
   P_n  \rightarrow P_n \langle \psi_n | \psi_n \rangle . 
\end{equation} 

\newpage
v) Normalize the $P_n$: $P_n  \rightarrow P_n  \left/  \sum_{n=1}^N P_n .  \right. $

vi) Normalize each of the $|\psi_n\rangle$. 

\vspace{1mm}
vii) Every few iterations perform the following operation (which might be referred to as ``splitting", ``breeding", or ``regenerating" the ensemble): For each pure state whose probability $P_j$ is less than a fixed threshold $P_{\mbox{\scriptsize thresh}} \ll 1$, we pick the state from the ensemble, $|\psi_m\rangle$, whose probability, $P_m$, is currently the largest in the ensemble. We then set $|\psi_j\rangle$ equal to $|\psi_m\rangle$, thus erasing $|\psi_j\rangle$ from the ensemble. We set both $P_j$ and $P_m$ equal to $P_m/2$.  Thus we have ``split" the highest probability state into two members of the ensemble, and this state is (most likely) no longer the highest contributing member.  After we have done this for each $P_j < P_{\mbox{\scriptsize thresh}}$, we then normalize all the $P_n$ as per v) above. 

3. The density matrix at time $t$ is (approximately) 
\begin{equation}
  \rho(t) =  \sum_{n=1}^N P_n(t) |\psi_n(t)\rangle \langle\psi_n(t) | . 
\end{equation} 

\subsection{Considerations for Numerical Accuracy} 

In the standard Monte Carlo method the only parameter that we must chose to reach a desired accuracy is $N$; we merely increase $N$ until we obtain this accuracy. For the new Monte Carlo method we have \textit{two} parameters that affect the error. The first is the minimum effective ensemble size during the evolution, $\mbox{min}(N_{\mbox{\scriptsize eff}})$. The second comes from the regeneration step. In each regeneration we eliminate some states. If we denote sum of the probabilities for these ``dropped" states as $P_{\mbox{\scriptsize drop}}$, then the maximum value of $P_{\mbox{\scriptsize drop}}$ during the simulation bounds the error from the regeneration step. So to ensure numerical accuracy we require that   
\begin{eqnarray}
     \mbox{min}(N_{\mbox{\scriptsize eff}}) & \gg & 1 ,   \nonumber \\ 
     \mbox{max}(P_{\mbox{\scriptsize drop}}) & \ll & 1 . 
\end{eqnarray} 
The values of these two quantities are determined jointly by $N$ and $P_{\mbox{\scriptsize thresh}}$. For a given value of $N$, there is some optimal value of $P_{\mbox{\scriptsize thresh}}$ that ensures that $\mbox{min}(N_{\mbox{\scriptsize eff}})$ is large while keeping $\mbox{max}(P_{\mbox{\scriptsize drop}})$ small. 

For a given simulation it is simple to check whether $N$ and $P_{\mbox{\scriptsize thresh}}$ give sufficient accuracy. One merely runs the simulation a second time with the same realization for the measurement noise $dW$, and different set of realizations for the noises that model the decoherence, $dV_i$. The difference between the two simulations gives one an estimate of the error. 

\subsection{Multiple Decoherence Channels and Multiple Measurements} 

For simplicity we presented the Monte Carlo method for an SME with only a single source of decoherence and single measurement.  Extending this to $m$ sources of decoherence and $l$ measurements is very simple. A system subjected to $l$ continuous measurements and $m$ sources of decoherence is described by the SME  
\begin{eqnarray}
 d\rho & = & - \sum_{i=1}^m \gamma ( L_i^\dagger L_i \rho + \rho L_i^\dagger L_i - 2 L_i \rho L_i^\dagger )dt \label{SMEm} \\ 
  & & - \sum_{j=1}^l k_j ( M_j^\dagger M_j \rho + \rho M_j^\dagger M_j - 2 M_j \rho M_j^\dagger )dt \nonumber \\ 
  & & + \sum_{j=1}^l \sqrt{2k_j}(M_j\rho +\rho M_j^\dagger - \langle M_j + M_j^\dagger \rangle \rho )dW_j \nonumber , 
\end{eqnarray}
where the $dW_j$ are mutually independent Wiener processes. To simulate this SME one simply repeats steps 2.\ i and ii for each of the $m$ decoherence channels, and steps 2.\ iii - vi for each of the $l$ measurement channels.  

\subsection{Inefficient Measurements} 

The form of the SME given in Eq.(\ref{SMEm}) above is general enough to include inefficient measurements~\cite{JacobsSteck06}. To make the $j^{\mbox{\scriptsize \textit{th}}}$ measurement inefficient we simply choose one of the $L_i$ to be equal to $M_j$,  and adjust the values of the corresponding $\gamma_i$ and $k_j$ to obtain the desired efficiency. 

\subsection{Using Milstien's Method for Time-Stepping} 

If one simulates a stochastic differential equation (SDE) simply by replacing $dt$ with a small time-step $\Delta t$, and $dW$ by a zero mean Gaussian random variable with variance $\Delta t$ (which we will call $\Delta W$), then the solution is only guaranteed to be accurate to half-order in $\Delta t$. Ensuring that the simulation is accurate to first-order in $\Delta t$ is simple, and the method for doing this is called \textit{Miltstien's} method. Milstein's method involves adding a term to the differential equation that is proportional to $(\Delta W^2 - \Delta t)$. The exact form of the Milstien term depends on the form of the stochastic term in the SDE. If the stochastic term is simply a linear operation, the this term is given by applying the linear operation twice, and multiplying by one half~\cite{Jacobs_SP}. Thus, the Milstien term for an SDE with the stochastic term $\alpha X |\psi\rangle$, for a number $\alpha$ and operator $X$, is 
\begin{equation}
  \Delta |\psi\rangle_{\mbox{\scriptsize Mil}} = \frac{\alpha^2}{2}(\Delta W^2 - \Delta t) X^2 |\psi\rangle . 
\end{equation}

\section{Deriving the method} 
\label{sec::deriv}

We begin by noting that if we apply the part of the evolution containing $L$ first, and that containing $M$ second, we get the evolution correct to first-order in $dt$. If the density matrix is given by $\rho = \sum_n P_n |\psi_n \rangle$, then the $L$ part of the evolution is obtained by using the standard Monte Carlo method (steps 2.\ i and ii above). To simulate the part containing $M$ (the measurement part), we note that this evolution can be written as~\cite{JacobsSteck06} 
\begin{equation}
  \rho(t + dt) =  \frac{1}{\mathcal{N}} A(\alpha) \rho(t) A^\dagger(\alpha) , \label{eqA} 
\end{equation} 
where $A$ is an operator that depends on the measurement result, $\alpha$, and $\mathcal{N}$ is simply an overall normalization factor. The measurement result $\alpha$ is the real number 
\begin{equation}
  \alpha =  2 k \langle M + M^\dagger \rangle dt + \sqrt{2\gamma} dW . 
\end{equation} 
By substituting  $\rho = \sum_n P_n |\psi_n \rangle \langle \psi_n |$ into Eq.(\ref{eqA}), we find that 
\begin{eqnarray}
     \rho(t + dt) & = & \frac{1}{\mathcal{N}} \sum_{n=1}^N P_n A |\psi_n \rangle \langle \psi_n | A^\dagger \nonumber \\ 
     & = & \sum_{n=1}^N \frac{P_n  \langle \psi_n | A^\dagger A |\psi_n \rangle}{\mathcal{N}}  \left[ \frac{A |\psi_n \rangle \langle \psi_n | A^\dagger }{\langle \psi_n | A^\dagger A |\psi_n \rangle}  \right] \nonumber \\ 
     & = & \sum_{n=1}^N P_n(t+dt) |\psi_n (t+dt) \rangle \langle \psi_n (t+dt) | , \nonumber 
\end{eqnarray} 
which gives us the simple update rules  
\begin{eqnarray}
     P_n(t + dt) & = & \frac{P_n(t) \langle \psi_n (t) | A^\dagger A |\psi_n (t)\rangle}{\mathcal{N}}  \\ 
     |\psi_n (t+dt) \rangle & = & \frac{A |\psi_n \rangle }{\sqrt{\langle \psi_n | A^\dagger A |\psi_n \rangle}} , 
\end{eqnarray} 
where $N$ is chosen so that $\sum_n P_n(t+dt) = 1$. From Eq.(29) in reference~\cite{JacobsSteck06}, the operator $A$ is 
\begin{eqnarray*}
     A(\alpha) =  1 - \gamma  \left[ M^\dagger - 2\langle M + M^\dagger \rangle \right] M dt +\sqrt{2\gamma} M  dW , 
\end{eqnarray*} 
and this gives us the evolution sequence for the Monte Carlo method presented above. 

The effect of the measurement is to increase the probabilities of some states, and reduce those of others. This reduces the effective size of the ensemble, and before too long there will only be one state left in the ensemble. To correctly model the noise being introduced into the system by the decoherence (the part of the evolution containing $L$) we need to maintain a large number of states in the ensemble. We solve this problem by using the ``regeneration" procedure (step 2.\ vii). Once every so-often we discard those states from the ensemble whose probabilities, and thus contribution, has become negligible. This discarding process does not effect the density matrix unduly so long as the total amount of probability of the discarded states is very small. Once the ``small" states have been discarded, we must choose new states to replace them, and we must do this without affecting the density matrix. This is easily achieved by duplicating some of the states that have a large contribution, and dividing the probability for each of these states equally between the original state and its duplicate. One should duplicate the states with highest probability, as this provides the biggest increase in the effective size of the ensemble. With the addition of this regeneration procedure, our Monte Carlo method is complete. 

\section{Example of a Numerical Simulation}
\label{sec::num}

Here we simulate a continuous measurement of the energy of a harmonic oscillator, using both the SME and the Monte Carlo method. We subject the oscillator to a randomly fluctuating white noise force, which serves as a simple model of (infinite temperature) thermal noise. The evolution due to the fluctuating force is given by~\cite{Jacobs_QMT} 
\begin{equation}
   d |\psi \rangle = \left[ - \frac{\beta}{2} x^2 dt + i \sqrt{\beta} x dW  \right] |\psi\rangle , 
\end{equation}
where $\beta$ determines the strength of the force. The term in this equation proportional to $dt$ is due to the transformation from Stratonovich to Ito noise. Since the observer does not know the fluctuating value of the force, she must average over it. The evolution of the observer's density matrix is then given by the master equation 
\begin{equation}
   d \rho  =  - \beta [x,[x,\rho]] dt .  
\end{equation}
Adding to this the evolution due the continuous measurement of energy (equivalently a continuous measurement of the phonon number, $N = a^\dagger a$), and the Hamiltonian evolution, the SME is 
\begin{eqnarray}
    d \rho & = &  - i\omega [N,\rho] dt - \frac{\beta}{2}  [x,[x,\rho]] dt - k [N, [N,\rho]] dt  \nonumber \\ 
    & & + \sqrt{2k}(N \rho + \rho N - 2 \langle N \rangle \rho) dW . 
\end{eqnarray}
Here $\omega$ is the frequency of the oscillator, $x = (a + a^\dagger)$ is the dimensionless position, and $k$ is the measurement strength. 

We simulate this equation using the unnormalized version 
\begin{eqnarray}
    d \rho & = &  - i\omega [N,\rho] dt - \frac{\beta}{2} [x,[x,\rho]] dt - k [N, [N,\rho]] dt  \nonumber \\ 
    & & + (N \rho + \rho N)(4k\langle N \rangle dt + \sqrt{2k} dW) \nonumber \\ 
    & & + k(dW^2 - dt)(N^2 \rho + \rho N^2 + 2 N \rho N) , 
    \label{eq::SMEsim}
\end{eqnarray}
and normalizing $\rho$ after each time-step. The reason we use this unnormalized version, which you will note is the same unnormalized version that we use for the wave-function in the Monte Carlo method, is that it makes the noise term in the SME linear in $\rho$, which in turn makes the Milstien term simpler to calculate. In the above equation the Milstien term is the final term. Since this equation is only accurate to first-order in the noise part of the evolution, we have also only included a first-order term for the deterministic evolution due to the Hamiltonian. 

The corresponding evolution for the Monte Carlo wave function is  
\begin{eqnarray}
    d |\psi\rangle & = &  - \left[  i\omega N - \frac{\beta}{2} x^2 - k N^2 \right] dt |\psi\rangle \\
    & & + (4k \langle N \rangle dt + \sqrt{2k} dW) N |\psi\rangle + i \sqrt{\beta} x dV |\psi\rangle \nonumber \\ 
    & & + k (dW^2 - dt) N^2 |\psi\rangle + \frac{\beta}{2} (dV^2 - dt)  x^2 |\psi\rangle \nonumber ,  
\end{eqnarray} 
where $dV$ is uncorrelated with $dW$.

\begin{figure}[t]
  \begin{center}
     \includegraphics[width=1\hsize]{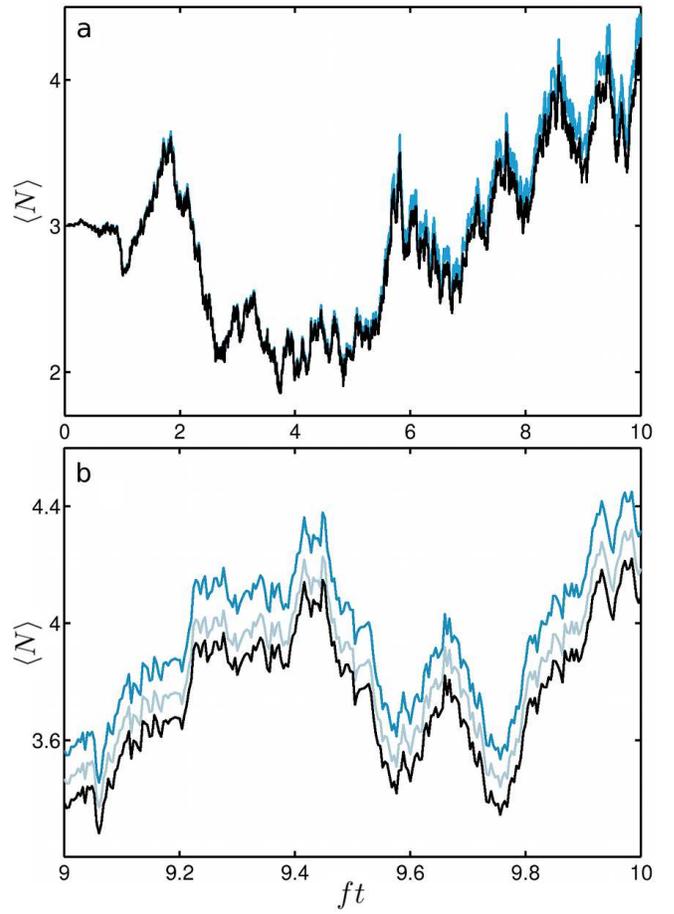}
  \end{center} 
  \caption
         {(Color online) The average value of the phonon number, $N$, for a continuous measurement of $N$, for an oscillator driven by a white-noise force. (a) A direct simulation of the SME (black line) and a simulation using the Monte Carlo method (grey line (blue online)). In this case the time-step is $dt = 2\times 10^{-4}T$, and the ensemble has $1024$ members. (b) A zoomed-in version of the Monte Carlo simulation in (a) (medium grey line (blue online)); a direct simulation of the SME with half the time-step (black line); and the MC simulation with half the time-step (light grey line). \label{fig1}}
\end{figure}

We now simulate the SME, using both Eq.(\ref{eq::SMEsim}) and the Monte Carlo method, and compare the results. For this simulation we set $k = g = 0.1f$ (where $f = \omega/2\pi \equiv 1/T$), start the oscillator in the Fock state with three phonons. We use a Fock-state basis, and truncate the state space at $9$ phonons. For our first run we choose $dt = 2\times 10^{-4} T$, and run for a time of $t = 10 T$.  For the Monte Carlo run we choose the ensemble size to be $N_{\mbox{\scriptsize ens}} = 1024$, and $P_{\mbox{\scriptsize thresh}} = 0.2/N_{\mbox{\scriptsize ens}}$. This choice results in $\mbox{min}(N_{\mbox{\scriptsize eff}}) = 745.2$ and $\mbox{max}(P_{\mbox{\scriptsize drop}}) = 0.003$. We plot the expectation value of the phonon number for both simulations in Fig.~\ref{fig1}a. We see that the results agree, but the solutions slowly diverge. To determine the source of this divergence we perform to more simulations. For the first one we double the size of the ensemble, and for the second we halve the time-step. Note that when we halve the time-step, we must use a noise realization that is consistent with that used for the first run, so that we can directly compare the trajectories in both cases~\cite{Jacobs_SP}.  

We find that doubling the size of the ensemble has little effect on the result of the Monte Carlo simulation. Halving the time-step, on the other hand, reduces the divergence between the two simulations considerably. In Fig.~\ref{fig1}b we plot the direct simulation of the SME using the smaller time-step, along with the Monte Carlo simulations using both time-steps. This plot is zoomed-in version of the trajectory in Fig~\ref{fig1}b. These results show us that the ensemble size of $1024$ is sufficient for this simulation, the inaccuracy being due almost entirely to the finite size of the time-step.  

\section{Conclusion}
\label{sec::conc} 
We have presented a wave-function Monte Carlo method for simulating systems that are under continuous observation, while also being subjected to noise and decoherence. This method is more efficient than the previously available method~\cite{Gambetta05}. We have also applied it to an example system, determining in this case sufficient resources to reproduce the SME. 

\textit{Note added:} Upon writing up this work, we discovered that a key element, that of ``splitting" the ensemble, had been introduced previously by Trivedi and Ceperley for Monte Carlo simulations of classical systems. See~\cite{Trivedi}. 

\section*{Acknowledgments} 
The author acknowledges the use of the supercomputing facilities in the school of Science and Mathematics at UMass Boston, as well as Prof.\ Daniel Steck's parallel cluster at the University of Oregon and the Oregon Center for Optics, which was funded by the National Science Foundation under Project No.\ PHY-0547926. This work was also supported by the National Science Foundation under Project No.\ PHY-0902906.


\end{document}